\documentclass[twocolumn,showpacs,preprintnumbers,amsmath,amssymb,aps,prl]{revtex4}
\usepackage{graphicx}
\begin{document}

\title{Rheology and Shear Band Suppression in Particle and Chain Mixtures }   
\author{I. Regev and C. Reichhardt } 
\affiliation{
Center for Nonlinear Studies and Theoretical Division,
Los Alamos National Laboratory, Los Alamos, New Mexico 87545 } 

\date{\today}
\begin{abstract}
Using numerical simulations, we consider an amorphous particle 
mixture which exhibits shear banding, and
find that the addition
of even a small fraction of chains 
strongly enhances the material strength,
creating pronounced overshoot features in the stress-strain curves.  
The strengthening occurs in the case where the chains 
are initially perpendicular to the shear direction, leading to 
a suppression of the shear band.
For large strain, the chains migrate to the region where a shear band forms, 
resulting in a stress drop.         
The alignment of the chains 
by the shear bands results in a Bauschinger-like effect 
for subsequent reversed shear. 
Many of these features are captured in a simple model of a 
single chain being pulled through a viscous material.
Our results are also useful for providing insights into methods of
controlling and strengthening granular materials against failure.
\end{abstract}
\maketitle

\vskip2pc
Sheared amorphous jammed materials such as colloids, foams, and 
granular media can exhibit a rich variety of dynamics,
such as strain localization (shear bands) \cite{Shall,Langer,J,G,Manning}, 
nonequilibrium phases \cite{Durian}, 
and dynamic ordering \cite{Daniels}. 
Most studies of these systems are performed with spherical or
elliptical particles; 
however, recent work has explored 
the packing or shearing of chains or granular polymer
systems \cite{Jaeger,Brown}.         
Jamming and packing studies of granular chains
indicate that as the chain length increases, the jamming density falls
further below 
the jamming density of monomer assemblies
\cite{Jaeger,L}.
In shear-strain experiments with three-dimensional (3D) granular chain
packings,
a strain stiffening behavior 
has been observed \cite{Brown} 
that resembles the behavior found in some polymer systems 
\cite{Hoy,Jones}. 
The granular chain experiments also revealed
a maximum in the stress as a 
function of strain that corresponds to the point at which the chains begin to
break. 
These results suggest that the addition of chains to monomer packings
could significantly alter the viscoplastic response, and that
granular materials could be strengthened by the addition of flexible
chains of some form.  
Understanding the effect of adding chains to a granular material could also have significant practical applications for preventing failures of granular piles or avalanches and for developing methods to
suppress shear band formation in materials. 

\begin{figure}[h]
\includegraphics[width=3.5in]{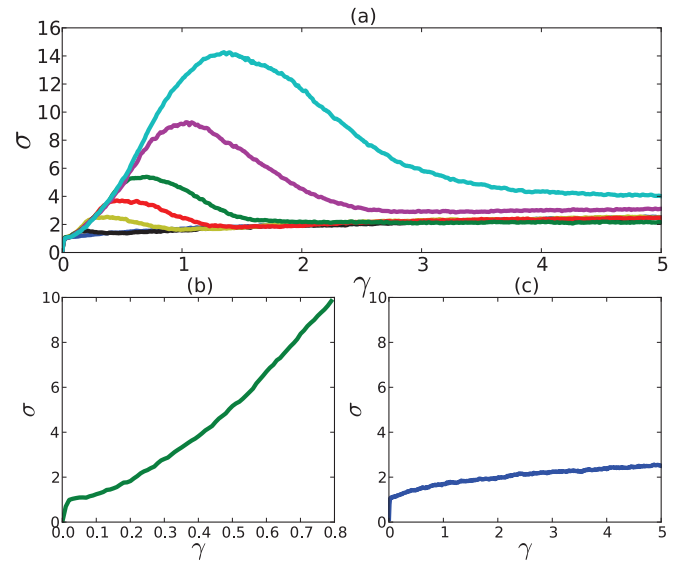}
\caption{
(Color online) (a) The stress-strain curves $\sigma$ vs $\gamma$ 
for an amorphous solid system with a chain added perpendicular to the
shear direction.  From bottom to top, chain length $L=0$ (no chain), 
$W/4$, $W/2$, $3W/4$, $W$, $3W/2$, and $2W$, where $W=\sqrt{N}$. 
Here ${\dot \gamma} = 5\cdot10^{-2}$, and the
chain-free system exhibits shear banding.  
A stress overshoot phenomenon develops as the chain length increases.
For $L = 2W$, the fraction of particles in the chain is $N_c/N=0.03125$. 
(b) A blowup of the $L=2W$ curve from (a)
showing strain stiffening.  (c) A blowup of the
chain-free system in (a) where strain stiffening is absent.
}
\label{fig:1}
\end{figure}

Here we examine the stress-strain
relations for an amorphous packing of particles that 
exhibits shear banding in the absence of chains. 
With the addition of even a small fraction of granular chains,
the system sustains remarkably higher amounts of 
stress and exhibits a stress-strain overshoot. 
The maximum stress at the peak of the overshoot increases with 
increasing chain length. The strengthening effect 
depends on the initial orientation of the chains with respect to the 
shear direction,
with the maximum effect occurring when the chains are initially perpendicular
to the shear. 
The effect arises due to the suppression by the chains 
of the formation of a shear band in the monomer component of the material.
Under increasing strain, the chains gradually migrate into the regions that
are prone to shear band formation,
leading to the stress decrease and the development of shear bands. 
After the peak in the stress overshoot, the stress returns to the value
found for a system containing only monomers.
The shear-induced
alignment of the chains also 
leads to a behavior that resembles the Bauschinger effect as the strain is cycled.
We show that many of the features of the system can be 
captured by a simple model of a chain 
immersed in a viscous fluid and pulled at one end.

{\it Simulation---} 
We consider a system of
$N$ particles with pairwise harmonic force interactions
${\bf F}^{i}_{dd} = \sum^{N}_{j\neq i}k_{g}(r^{ij}_{\rm eff} - R_{ij})\Theta(r^{ij}_{\rm eff} - R_{ij}){\bf {\hat R}}_{ij}$  
with a spring constant of $k_{g} = 300$, 
where ${\bf R}_{i(j)}$ is the position of particle $i(j)$,
$R_{ij} = |{\bf R}_{i} - {\bf R}_{j}|$,
${\bf {\hat R}}_{ij} = ({\bf R}_{i} - {\bf R}_{j})/R_{ij}$, 
$\Theta$ is the Heaviside step function,
$r^{ij}_{\rm eff} = r_{i} + r_{j}$, and $r_{i(j)}$ is the radius of disk $i(j)$. 
To ensure that
the disk packings are amorphous, we use a binary distribution 
of disks where we set half the disk radii to $r_{i} = 1.0$ 
and the other half to $r_{i} = 1.4$. The chains
are modeled as connected disks that 
have the same harmonic interactions with the surrounding disks and experience
an additional force
$F^{i}_{c} = \sum_{k}k_{c}k_{g}(r^{ik}_{\rm eff} - R_{ik}){\bf {\hat R}}_{ik}$
where the disks $k$ are the immediate neighbors of the disk $i$ 
along the chain.  Here $k_c$ is the spring constant of the chain
interaction, and a given chain contains $N_c$ particles.
We use an overdamped equation of
motion which gives athermal dissipative dynamics, 
$m_{i}{\bf {\dot p}}_i = {\bf F}^{i} - \eta{\bf v}_{i}$,
with ${\bf F}^i={\bf F}^i_{dd}+{\bf F}^i_c$.
Here the damping coefficient $\eta = 10$ and the masses of the disks 
are set to $m_{i} = 1.0$.  The equations
of motion are solved using the Leap-Frog algorithm. 
To initialize the system, we place one or more chains 
in a region of the plane, and randomly distribute other monomers around 
it. We then compress the system to a density
$\phi = 1.1624$; this is well into the jammed state, as the 
jamming density for this system has been shown to occur
at $\phi_{j} \approx 0.844$ \cite{Tietel,Cl,Nagel,Drocco}. 
Shear is induced via the Lees-Edwards boundary conditions \cite{72LE}.
These boundary conditions have previously been employed
for the chain-free version of this system near jamming \cite{Tietel,Cl}. Similar boundary conditions have been employed in experiments studying shear-bands \cite{03FV1, 03FV2}. 

{\it Results---}
In the absence of a chain, the monomer system exhibits
a shear banding effect when 
${\dot \gamma}$ is sufficiently large. 
Figure~1(a) shows the stress-strain $\sigma - \gamma$ curves for a 
system with $N = 4096$ disks with 
${\dot \gamma} = 5\cdot10^{-2}$, which is large enough for shear banding
to occur in the chain-free system. 
For a chain-free sample, there is a linear elastic regime followed by 
 yield and a plastic regime where a shear band forms, as illustrated
in Fig.~1(c).
At higher $\gamma$, 
$\sigma$ is roughly constant with a value near $\sigma=2.0$.
When a chain of length $L$ is placed in the system
perpendicular to the shear direction, 
we observe the development of a stress overshoot that becomes larger as
$L$ increases.  This is shown in Fig.~1(a) for chains of length ranging from
$L=W/4$ to $L=2W$, where $W=\sqrt{N}$ is the width of the system.
For the longest chain $L=2W$, which contains a fraction of only 
$N_c/N=0.03125$ (about $3\%$) of the total number of grains in the system,
the peak value of $\sigma$ 
is almost ten times 
higher than for the chain-free system. 
In Fig.~1(b), a blowup of the $\sigma-\gamma$ curve 
for the $L=2W$ system illustrates
a strain stiffing response, indicated by $\sigma$ 
growing nonlinearly with $\gamma$.  The initial increase of
$\sigma$ can be fit approximately  
to $\sigma \propto \gamma^{2}$. 
Figure 1(c)
shows the chain-free system where strain stiffening is absent (the system actually experiences mild strain-softening).   
These results are similar to those observed
in experimental studies of 3D systems of pure chains, 
where a strain stiffing response is most pronounced
for longer chains that also have a peak in the stress-strain 
curves \cite{Brown}. 
The strain stiffening was attributed in \cite{Brown} to entanglements, which
do not occur in our system due to the diluteness of our chains and the
two-dimensional (2D) nature of our system.
Additionally, in the experimental system, the drop in the stress above the 
peak response is due to the chains breaking, something that is not permitted to
occur in our system.
\begin{figure}
\includegraphics[width=3.5in]{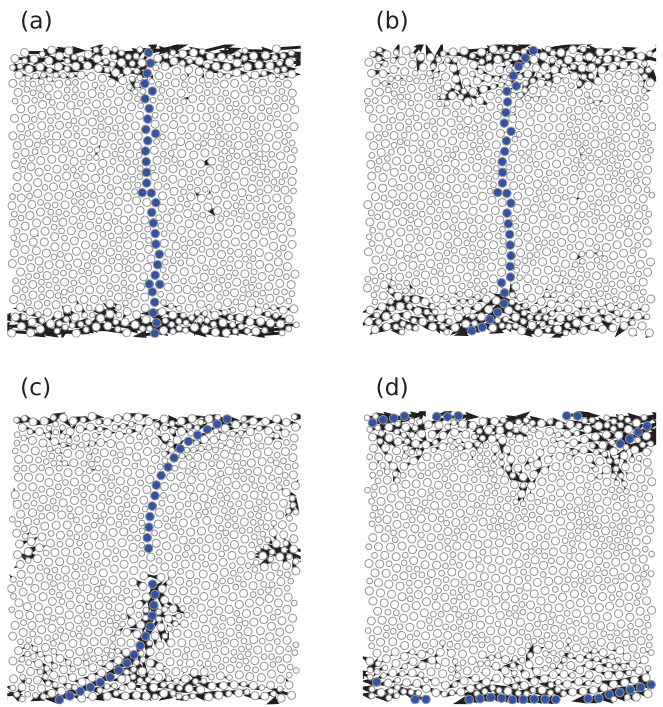}
\caption{ 
(Color online) 
Particle locations (open circles), chain location (filled circles), 
and motion (arrows) for the system in Fig.~1 with chain length $L = W$. 
(a) The initial state at $\gamma = 0.013$ where the chain is aligned 
perpendicular to the shear direction.
(b) $\gamma = 0.25$. 
(c) $\gamma = 0.625$, close to the peak of the stress.
(d) $\gamma = 2.0$, after the stress drop
(for a more vivid visualization see also animation in supplementary material).}   
\label{fig:diverge}
\end{figure}

A better understanding of the $\sigma-\gamma$ overshoot 
is provided by the images in Fig.~2 of a chain with $L = W$ immersed in a smaller system $N=1024$. 
In Fig.~2(a), taken at the very small strain of $\gamma=0.013$, 
the initial placement of the chain perpendicular to the shear direction
is barely disturbed.
As $\gamma$ increases to $\gamma=0.25$, shown in Fig.~2(b), 
the chains begin to stretch and form an S shape at
the top and bottom of the sample where the shear band 
forms in the absence of a chain. 
Figure 2(c) shows that at $\gamma=0.625$, where the peak value of $\sigma$
occurs, the chain has become fully stretched and is beginning to migrate
into the shear band region, 
while at $\gamma = 2.0$ beyond the peak stress, Fig.~2(d) indicates that
the chain has fully migrated into the shear band region and that
the shear band is fully formed.   
The peak in $\sigma$ occurs due to the
suppression of the shear band in the monomers 
by the creation by the chain 
of an effective coupling between the particles in the shear band region
and the particles in the bulk region at the center of the sample.
The shear band develops rapidly
once the chain aligns with the shear direction, 
leading to the stress drop. 
\begin{figure}
\includegraphics[width=3.5in]{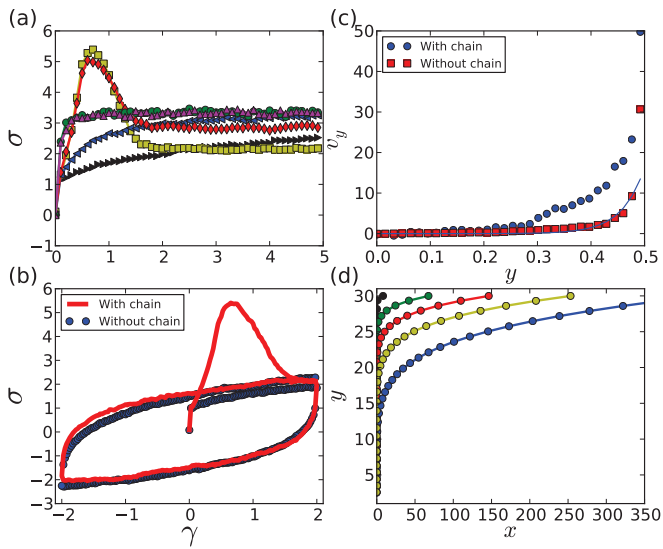}
\caption{
(Color online)
(a) The effect of strain rate on the $\sigma-\gamma$ curves 
for ${\dot \gamma} = 10^{-3}$, $10^{-2}$, and $5\times 10^{-2}$ for the
chain-free (green circles, blue left-triangles and black-right triangles respectively) and $L=W$ (magenta up-triangles, red diamonds and yellow squares respectively) samples.
At ${\dot \gamma} = 10^{-3}$ in the sample containing a chain,
there is no shear band formation
and the overshoot phenomenon is absent.  
(b) $\sigma-\gamma$ curves for the system in Fig.~1 
with $L=W$ (thin red curve) and with no chain (blue circles) obtained for
cyclic applied strain.
Here the stress overshoot only occurs
during the initial strain increase in the presence of the chain,
and the chain remains inside the shear band during subsequent
strain cycles.
(c) Velocity profile of the particles near the moving boundary at 
$\gamma=0.05$. The red squares are the curve for the system with no chain
that show the usual exponential decay (solid blue line is an exponential fit) 
as expected \cite{G, Shall}. 
The blue circles are the curve for the system 
with a chain that does not fit an exponential profile.
(d) Chain configuration at different times
for a simple model of a chain with $k_c=0.3$ in a viscous media pulled 
by the top bead with constant $v_x$.  Time increases from 
the left-most (black) curve to the right-most (blue) curve.
}
\label{fig:vt}
\end{figure}
A further indication that the realignment of the chain with the shear band
is responsible for the stress drop appears in Fig.~3(b), where we
plot the $\sigma-\gamma$ curve for the $L=W$ sample under cyclic applied
strain.

As was mentioned above, 
the hardening effect appears to be related to the suppression of shear-bands. 
Since shear bands in granular matter appear for high strain rates (see, 
for example, \cite{11CWBNS}), 
changing the strain rate $\dot\gamma$ should alter the hardening behavior. 
In Fig.~3(a) we plot the $\sigma-\gamma$
curves for a chain-free system at ${\dot\gamma} = 10^{-3} , 10^{-2}$ and
$5\times 10^{-2}$ along with
the corresponding curves in the presence of an $L=W$ chain. 
In the chain-free system, the steady state value of $\sigma$ for
${\dot \gamma}=10^{-3}$ where shear banding does not occur
is higher than for samples with higher ${\dot \gamma}$ where shear
banding takes place.
When an $L=W$ chain is added to the system,
at ${\dot \gamma} = 10^{-3}$ the $\sigma-\gamma$ curve is almost 
identical to that found for the chain-free case. 
Although the chain still gradually aligns with the shear direction, there
is no shear band to block at this strain rate, and thus no difference in
response when the chain is added. The hardening and stress overshoot due to the presence of a chain appears only at rates high enough for shear band formation, 
shown in Fig.~3(a).
At large strains, the samples containing a chain have a lower value of 
$\sigma$ than the chain-free systems subjected to the same strain and
strain rate.
This is due to a lubrication effect which also seems to be important only at higher strain-rates. 
This indicates that in some cases the addition of a chain can decrease
rather than increasing the stress.

An overshoot only occurs during the
initial increase in the strain when the chain has not yet aligned with the
shear band.
Once the chain has entered the shear band region, it remains there even when 
the strain rate is reversed, and subsequent cycles of the strain produce stress curves
that closely follow those obtained for a system without chains, as shown
in Fig.~3(b).
The change in the $\sigma-\gamma$ curves upon reversing the strain
resembles the Bauschinger effect - the material responds differently to positive or negative shear on the same axis - which occurs in amorphous solids and crystals. In the chain system this effect vanishes after the first shearing cycle. In Fig.~3(c) we can observe the difference in the velocity profile between the chain-free and chain configurations. The chain-free configuration shows the typical exponential behavior (see \cite{G, Shall}) while the chain configuration shows a more gradual velocity profile.

To examine the effects of the chain stiffness, 
in Fig.~4(a) we plot 
the $\sigma-\gamma$ curves for a sample containing a chain with
$L = W$ at ${\dot \gamma}= 5\times 10^{-2}$ 
for $k_{c} = 2.5$, 5, 10, and 20.  The stress peak increases 
with decreasing $k_{c}$ and the position of the peak shifts to higher strain. 
Figure~4(b) shows the fraction $f_{\rm perp}$ of the chain length oriented in
the direction perpendicular to the strain versus $\gamma$.
The chain transitions from being aligned completely perpendicularly
to the strain ($f_{\rm perp}=1$) to being aligned with the
shear direction ($f_{\rm perp}=0$).
The softer chain is more difficult to pull through the bulk due to its
greater tendency to meander as it interacts with the monomers.
We have also tested multiple chains in the same system and found
an additive effect where $\sigma$ in a sample containing two $L=W$
chains is twice as large as $\sigma$ in a sample
containing only one $L=W$ chain. 

For very long chains, we observe the same general behaviors described above;
however, additional nonlinear rheological effects can arise that we will
describe in greater detail in another paper.
Many of the features we observe in the simulation can be captured 
by a simple model of a single chain pulled at one end through a viscous 
medium. The model consists of $L$ point particles with 
mass $m =1$ connected by springs with constant spring force $k_c$ 
and a rest length $r_{0} = 1$. Each particle experiences
a drag force $f^i_{drag}$ that is assumed to depend on the vertical distance 
$\delta y$
between the particle and its lower neighbor as well as on the
horizontal velocity of the particle: $f^i_{drag}\propto\delta y_{i,i-1}v^i_x$.
The particles also experience a uniform damping force
$f_{damp}=\eta v_i$.
We fix the top of the chain to be at a constant height 
$y = W$ and to move at constant velocity $v_{x}$.
This represents the case where the top of the chain is dragged by a
shear band and pulled at a fixed rate.
In Fig.~3(d) we plot the chain configurations from this model at different
times, showing that the chain gradually moves to flow behind the driven
bead in a manner similar to that observed in our simulations in Fig.~2.
In Fig.~4(c) we plot $F_{\rm tot}$ versus time for chains of length $L=30$ 
and
increasing $k_c$.  The peak value of $F_{\rm tot}$ increases with decreasing
$k_c$, similar to the behavior found in our simulation.
Fig.~4(d) shows the total force $F_{\rm tot}$ exerted on the chain by the
driven bead versus time for increasing chain length $L$. 
A peak in $F_{\rm tot}$ occurs and is followed
by a drop off, with the peak shifting to later strains and increasing in height
as $L$ increases.
There are several differences between the model and the simulation,
including the fact that the strain stiffening effect does not appear in the
model when the strain is initially increased, and the magnitude of the
peak in $F_{\rm tot}$ does not increase as rapidly with $L$ in the model as 
does the peak in $\sigma$ in the simulation.
The model does, however, capture most features of the shapes of the
curves and the behavior of the chain.
\begin{figure}
\includegraphics[width=3.5in]{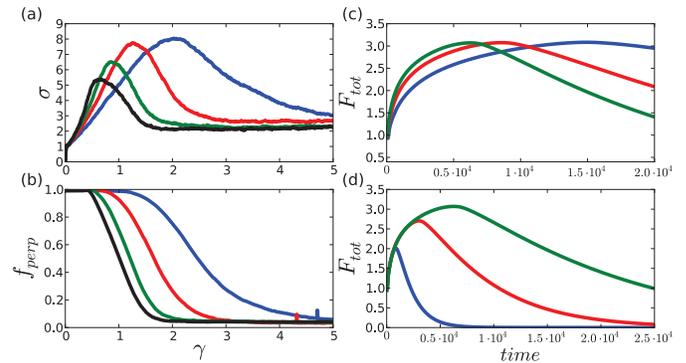}
\caption{ 
(Color online)
(a) $\sigma - \gamma$ curves for the system in Fig.~1 with
$L = W$ and ${\dot \gamma} = 5\times 10^{-3}$ for varied 
chain spring constant $k_{c}=2.5 ,5,20$ and $20$  where weaker chains 
produce larger but less stiff peaks.
(b) The corresponding fraction $f_{\rm perp}$ 
of the chain oriented in the direction
perpendicular to the strain.
This figure shows that the softer chains generate a higher stress peak due to 
spending more time in the bulk before being pulled to the shear-band.
(c) The total force $F_{\rm tot}$ applied to the chain in the simple model
vs time for $L = 30$ with $k_{c}=0.3$, $0.2$, and $0.1$, from left to right, 
showing the increase in stress with decreasing $k_{c}$.   
(d) 
$F_{\rm tot}$ vs time in the simple model
for increasing chain length $L$ showing an overshoot and 
an increase in the $F_{\rm tot}$ for longer chains. 
From left to right, $L=10$, $20$, $30$.  Here $k_c=0.3$.
}
\label{fig:fraction}
\end{figure}

In summary, we have shown that the addition of a small fraction of chains
to an amorphous solid can strongly alter the stress-strain response and
can significantly strengthen the material, producing a pronounced overshoot
in the stress-strain curve.
This effect occurs for strain rates 
at which the chain-free system exhibits shear 
banding.  
When the chains are initially oriented perpendicular to the strain direction, 
the shear banding is partially suppressed until the chain 
becomes aligned with the shear direction and absorbed into the shear band,
leading to a stress drop. 
The alignment of the chains into the
shear band also produces a Bauschinger like effect for cyclic strains.
For long chains we also observe a strain stiffing
effect that is absent in the chain-free samples. 
Our results have several similarities to recent experiments in 
3D systems of pure granular chains, including strain
stiffing and stress overshoots; however, there are significant differences, 
including the fact that the stress drop after the overshoot
arises due to chain alignment rather than chain breaking, and the fact that
there are no entanglements in our system. 
Our results suggest that the response of granular media can be significantly 
altered and the media strengthened by the addition of even a small number of 
chain like particles.  This could have applications for preventing failures 
of granular heaps and soils or preventing erosion and is another example of the importance of structure to the rheology of amorphous materials \cite{10BHPRZ, 98FL}. 

The authors thank Lena Lopatina for useful discussions 
and Cynthia Reichhardt for carefully reviewing the manuscript.
This work was carried out under the auspices of the 
NNSA of the 
U.S. DoE
at 
LANL
under Contract No.
DE-AC52-06NA25396.

\end{document}